\documentclass{ifacconf}

\usepackage{graphicx}      
\usepackage{natbib}  

\usepackage{tikz}
\usepackage{dsfont}
\usepackage{amsmath}
\usetikzlibrary{angles}

\usepackage{amsmath} 
\usepackage{amsfonts}

\usepackage{float} 
\usepackage[skip=0em,font=footnotesize]{caption}

\usepackage{tikz}
\usetikzlibrary{shapes,arrows,calc,positioning}
\tikzstyle{bigblock} = [draw, fill=blue!20, rectangle, 
    minimum height=6em, minimum width=8em]
\tikzstyle{medblock} = [draw, fill=blue!20, rectangle, 
    minimum height=4em, minimum width=4em]    
\tikzstyle{mux} = [draw, fill=black!20, rectangle, 
    minimum height=5em, minimum width=0.1em]    
\tikzstyle{smallblock} = [draw, fill=blue!20, rectangle, 
    minimum height=2em, minimum width=3em]
    
\tikzstyle{data_block} = [draw, fill=green!20, rectangle, 
    minimum height=2em, minimum width=3em]
\tikzstyle{ops_block} = [draw, fill=blue!20, rectangle, 
    minimum height=2em, minimum width=3em]    
\tikzstyle{est_block} = [draw, fill=red!20, rectangle, 
    minimum height=2em, minimum width=3em]    
    
\tikzstyle{sum} = [draw, fill=blue!20, circle, node distance=1cm,minimum height=0.5cm]
\tikzstyle{signal} = [coordinate]
\tikzstyle{pinstyle} = [pin edge={to-,thin,black}]
\tikzstyle{block} = [draw, fill=blue!20, rectangle, 
    minimum height=3em, minimum width=9em]
\tikzstyle{blockS} = [draw, fill=blue!20, rectangle, 
    minimum height=3em, minimum width=4em]    
\tikzstyle{input} = [coordinate]
\tikzstyle{output} = [coordinate]
\usetikzlibrary{matrix}

\usetikzlibrary{positioning}
\usetikzlibrary{math}

\usepackage{xcolor}

\newcommand{\bc}{\begin{center}}
\newcommand{\ec}{\end{center}}
\newcommand{\benum}{\begin{enumerate}}
\newcommand{\eenum}{\end{enumerate}}
\newcommand{\nn}{\nonumber}
\newcommand{\matl}{\left[ \begin{array}}
\newcommand{\matr}{\end{array} \right]}

\renewcommand{\matl}{\begin{bmatrix}}
\renewcommand{\matr}{\end{bmatrix}}

\newcommand{\matls}{\left[ \begin{smallmatrix}}
\newcommand{\matrs}{\end{smallmatrix} \right]}
\newcommand{\isdef}{\stackrel{\triangle}{=}}

\newcommand{\rmA}{{\rm A}}
\newcommand{\rmB}{{\rm B}}
\newcommand{\rmC}{{\rm C}}
\newcommand{\rmD}{{\rm D}}
\newcommand{\rmE}{{\rm E}}

\newcommand{\rmG}{{\rm G}}

\newcommand{\rmI}{{\rm I}}

\newcommand{\rmK}{{\rm K}}

\newcommand{\rmM}{{\rm M}}

\newcommand{\rmR}{{\rm R}}
\newcommand{\rmS}{{\rm S}}
\newcommand{\rmT}{{\rm T}}

\newcommand{\rmW}{{\rm W}}

\newcommand{\rmc}{{\rm c}}

\newcommand{\rmf}{{\rm f}}

\newcommand{\rmi}{{\rm i}}

\newcommand{\rms}{{\rm s}}

\newcommand{\rmu}{{\rm u}}

\newcommand{\rmw}{{\rm w}}
\newcommand{\rmx}{{\rm x}}

\newcommand{\BBR}{{\mathbb R}}

\usepackage{enumitem,amssymb}
\newlist{todolist}{itemize}{2}
\setlist[todolist]{label=$\square$}
\usepackage{pifont}
\begin{document}

\begin{frontmatter}
%
\title{\LARGE Data-driven Fuzzy Control for Time-Optimal Aggressive Trajectory Following}

\author[First]{August Phelps}
\author[First]{Juan Augusto Paredes Salazar}
\author[First]{Ankit Goel}


\address[First]{Department of Mechanical Engineering, \\ University of Maryland, Baltimore County, \\ 1000 Hilltop Circle, Baltimore, MD 21250.  \\ (e-mail: \{aphelps1,japarede,ankgoel\}@umbc.edu).}

\begin{abstract}

Optimal trajectories that minimize a user-defined cost function in dynamic systems require the solution of a two-point boundary value problem.
The optimization process yields an optimal control sequence that depends on the initial conditions and system parameters. 
However, the optimal sequence may result in undesirable behavior if the system's initial conditions and parameters are erroneous. 
This work presents a data-driven fuzzy controller synthesis framework that is guided by a time-optimal trajectory for multicopter tracking problems.
In particular, we consider an aggressive maneuver consisting of a mid-air flip and generate a time-optimal trajectory by numerically solving the two-point boundary value problem. 
A fuzzy controller consisting of a stabilizing controller near hover conditions and an autoregressive moving average (ARMA) controller, trained to mimic the time-optimal aggressive trajectory, is constructed using the Takagi-Sugeno fuzzy framework. 

\end{abstract}



\end{frontmatter}

\maketitle


\section{Introduction} \label{sec:introduction}
%
%
Trajectory optimization in dynamic systems involves the minimization of a user-defined cost function while satisfying dynamic constraints, imposed by the system dynamics, as well as user-defined state and control constraints. 
Given the dynamic nature of the constraints, trajectory optimization leads to a two-point boundary value problem, which is typically challenging to solve, even in straightforward scenarios.
In practice, the optimal trajectory problem is solved numerically and produces an optimal sequence of states and controls \cite{gasparetto2015}. 
Several numerical solvers exist to solve such optimal control problems, such as DIDO \cite{ross2020enhancements}, GPOPS \cite{rao2010algorithm, patterson2014gpops}, and CasADi \cite{andersson2019}.
%
However, the optimal control sequence cannot be applied directly to the system, as any perturbation in initial conditions or physical properties of the system leads to divergence from the optimal state sequence. 
Furthermore, recomputing the optimal control sequence along the trajectory is also not feasible due to high computational cost of the numerical solvers. 


%
To follow the optimal trajectory, a feedback controller can be used to stabilize the state error dynamics. 
However, designing feedback controllers to stabilize the error dynamics can require substantial computational resources if the system dynamics are complex \cite{nguyen2020, shadeed2021, li2024, wang2021}.
%
%
%
%

%
This paper is therefore focused on designing a computationally inexpensive feedback controller to track an optimal trajectory. 
In particular, we consider the problem of tracking a time-optimal aggressive maneuver that involves a multicopter flip and constraints. 
A time-optimal trajectory is first computed using the CasADi solver. 
Next, a linearization-based fullstate feedback (LBFSF) controller is designed to follow the optimal state trajectory in the case of perturbations from the optimal state trajectory. 
Next, to reduce the computational cost of implementing the LBFSF controller, we train an autoregressive moving average (ARMA) controller with the input-output data generated with the LBFSF controller. 
Finally, a Takagi-Sugeno (T-S) fuzzy system \cite[ch.~6]{lilly2011}  is designed to improve the robustness of the ARMA controller.
The T-S fuzzy framework is chosen since it provides an intuitive methodology to interpolate the response of linear systems for control applications \cite{nguyen2019,al2023,precup2024}.
In this paper, we consider a bicopter system, which is a multicopter that is constrained to move in a vertical plane, since the bicopter retains the nonlinear characteristics of the multicopter but requires reduced computational resources.

The contents of this paper are as follows.
Section \ref{sec:bicopter_dyn} reviews the dynamics of the bicopter considered in this work.
Section \ref{sec:control} reviews the optimal trajectory generation problem and the feedback controllers. 
%
%
%
%
%
Section \ref{sec:numerical_examples} presents a simulation results to demonstrate the application of the proposed technique to track time-optimal aggressive trajectories using a feedback controller. 
%
%
%
Finally, the paper concludes with a summary in Section \ref{sec:conclusions}.

\section{Bicopter Dynamics}\label{sec:bicopter_dyn}

\begin{figure}[h]
    \centering
    \includegraphics[width=0.6\columnwidth]{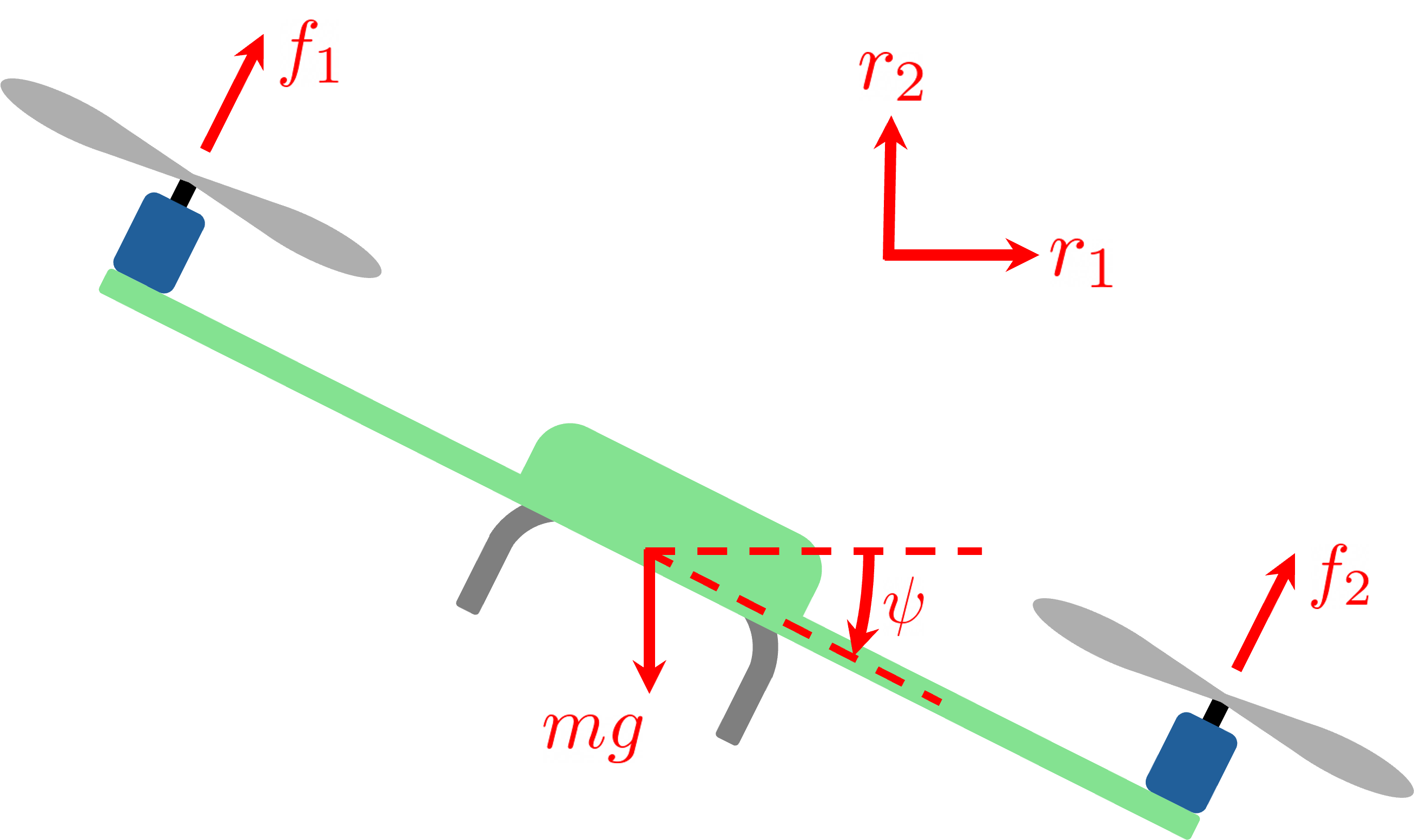}
    \caption{A bicopter.}
    \label{fig:quad_model}
\end{figure}

We consider the bicopter shown in Figure \ref{fig:quad_model}.
Let $r_1,r_2,\psi \in \BBR$ be the horizontal position, height, and pitch angle of the bicopter, respectively.
Let $m$ and $J$ denote the mass and inertia of the bicopter. 
Let $f_\rmT$ denote the total force and $\tau$ denote the total moment applied to the bicopter. 
As detailed in \cite{delgado2024adaptive}, the equations of motion are
%
%
\begin{align}
    \ddot{r}_1 = \frac{f_\rmT}{m} \sin\psi, \quad
    \ddot{r}_2 = \frac{f_\rmT}{m} \cos\psi - g, \quad
    \ddot{\psi} = \frac{\tau}{J}, \label{eq:dyn_theta}
\end{align}
where $g = 9.81$ $\rm  m/s^2 $ is the acceleration due to gravity.
%

Defining the state 
\begin{equation}
   x \isdef \matl r_1 & \dot{r}_1 & r_2 & \dot{r}_2 & \psi & \dot{\psi} \matr^\rmT \in \mathbb{R}^6,
\end{equation}
and the input 
\begin{equation}
    u \isdef \matl u_{\rm T} & u_{\rm R} \matr^\rmT \in \mathbb{R}^2,
\end{equation}
where $u_\rmT \isdef f_\rmT / m$ is the normalized thrust and $u_\rmR \isdef \tau / J$ is the normalized torque, the system dynamics \eqref{eq:dyn_theta} can be written as
\begin{equation}\label{mult_dyn}
    \dot{x} = f(x, u) = \begin{bmatrix} \dot{r}_1 \\ u_{\rm T} \sin \psi \\ \dot{r}_2 \\ u_{\rm T} \cos \psi - g \\ \dot{\psi} \\ u_{\rm R} \end{bmatrix}.
\end{equation}


\section{Control} \label{sec:control}
This section reviews the time-optimal trajectory generation algorithm, feedback controllers considered in this work, and the fuzzy system to combine the feedback controllers. 
In particular, a nonlinear program to minimize the time-of-flight while satisfying user-specified state and control constraints is described. 
Next, a linearization-based fullstate feedback controller and an autoregressive moving average controller trained with input-output data are described. 
Finally, a fuzzy system to combine both of the feedback controllers is described. 

\subsection{ Time-Optimal Trajectory Generation} \label{sec:NLP}

Consider the system
\begin{align}
    \dot{x} (t) &= f(x(t),u(t)), \label{eq:NL_system_state}
\end{align}
where
$x(t) \in \BBR^{\ell_x}$ is the state of the system, 
$u(t) \in \BBR^{\ell_u}$ is the input applied to the system, and
$y(t) \in \BBR^{\ell_x}$ is the output of the system.
The function $f \colon \BBR^{\ell_x} \times \BBR^{\ell_u} \to \BBR^{\ell_x}$ encodes the system dynamics.
%
%
Discretizing \eqref{eq:NL_system_state} with Euler's method and time step $T_{\rm dt}$ yields
\begin{align}
    x_{k+1} = x_k + T_{\rm dt} f(x_k, u_k), \label{eq:NL_system_state_DT}
\end{align}
where $k\ge0$ is the iteration step.
Note that continuous time $t$ and the discrete-time step $k$ are related by $t = k T_{\rm dt}.$
%

%
The minimum-time trajectory optimization objective is to drive the state $x_k$ from an initial state $x_\rmi$ to a desired state $x_\rmf$ while ensuring that both the state $x_k$ and the control $u_k$ at each time step $k$ satisfy user-specified constraints in minimum time.
%
%
Furthermore, consider a desired final input $u_\rmf.$
This objective is mathematically equivalent to the nonlinear program
%
%
%
\small
\begin{align}
    &\min_{(x_k)_{k=0}^N, (u_k)_{k=0}^{N-1}, T} T +  (x_\rmf - x_N)^\rmT Q_{x} (x_\rmf - x_N) \nn \\ 
    & \hspace{8em} + (u_\rmf - u_N)^\rmT Q_{ u} (u_\rmf - u_N), \label{eq:NLP_init}
\end{align}
\normalsize
subject to
\small
\begin{align}
    &x_{k+1} = x_k + T_{\rm dt} f(x_k, u_k),  \mbox{ for all } k\in\{0, \ldots, N-1\},\\
    &x_0 = x_\rmi, \quad x_N = x_\rmf, \quad u_N = u_\rmf,\label{eq:NLP_Constraints}\\
    &x_{k} \in [x_{{\rm min}}, x_{{\rm max}}], \\
    %
    &u_{k} \in [u_{{\rm min}}, u_{{\rm max}}], 
    \\
    &T > 0, \label{eq:NLP_fin}
\end{align}
\normalsize
where $T$ is the time of flight that is minimized, and the positive semidefinite matrices $Q_x \in \BBR^{\ell_\rmx \times \ell_\rmx}$  and  $Q_u \in \BBR^{\ell_\rmu \times \ell_\rmu}$ are the final state error and the final input error weighting matrices, respectively.
%

%
Note that the optimization variables of the NLP given by \eqref{eq:NLP_init}--\eqref{eq:NLP_fin} are the final time $T,$ the discrete-time trajectory states $x_0, x_1 \ldots, x_N,$ and the discrete-time input states $u_0, u_1 \ldots, u_{N-1}.$
The number of discrete-time iterations $N$ is chosen by the user and is fixed throughout the optimization procedure, which implies that $T$ determines the time between iterations $T_{\rm dt} = T/N.$
Hence, increasing $N$ may yield smaller values of $T_{\rm dt},$ which may allow smoother optimal trajectories to be obtained and increase the feasibility of solving the NLP.
The resulting optimal final time is given by $T^*,$ and, for all $k \in \{0, \ldots, N\},$ the resulting optimal states and inputs are given by $x_k^*$ and $u_k^*,$ respectively.
%
%
In this work, we use the CasADi software toolbox \cite{andersson2019} to solve the nonlinear program described above.

\subsection{Feedback Control}
\label{sec:fbc}
Although the solution of NLP described in the previous section provided the optimal input and state sequence, any perturbation from the optimal solution leads to divergence since the system is inherently unstable. 
In this section, we design feedback controllers that stabilize the optimal trajectory and ensure that the bicopter state remains close to the optimal state trajectory. 
Specifically, we consider the control law
\begin{align}
    u(t) = u^*(t) + u_{\rm fb}(t),\label{eq:u_fb}
\end{align}
where $u_{\rm fb}(t) $ is the output of the feedback controller designed below.

\subsubsection{Linearization-based fullstate feedback (LBFSF) control along optimal trajectory} \label{sec:fbc_opt}
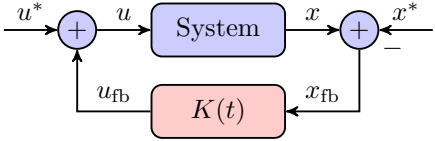
\begin{figure}[h]
\centering
%
{%
    \begin{tikzpicture}[>={stealth'}, line width = 0.25mm]

    \node [input, name=ref]{};
    \node [smallblock, rounded corners, right = 0.5em of ref , minimum height = 2em, minimum width = 5em] (system) {System};
    \node [sum, left = 2em of system] (sum1) {};
    \node[draw = none] at (sum1.center) {$+$};
    \node [sum, right = 2em of system] (sum2) {};
    \node[draw = none] at (sum2.center) {$+$};
    \node [smallblock, fill=red!20, rounded corners, below = 1 em of system, minimum height = 2em , minimum width = 5em] (controller) {$K(t)$};
    \draw [->] (controller.west) -| node [above, near start] {$u_{\rm fb}$} (sum1.south);
    \draw[->] (system.east) -- node [above]{$x$} (sum2.west);
    \draw[->] (sum2.south) |- node [above, near end] {$x_{\rm fb}$} (controller.east);
    \draw[->] (sum1.east) -- node [above] {$u$} (system.west);
    \draw[->] ([xshift = -2em]sum1.west) -- node [above] {$u^*$} (sum1.west);
    \draw[<-] (sum2.east) -- node [above] {$x^*$} node [below, xshift = -0.5em] {$-$} ([xshift = 2em]sum2.east);
    \end{tikzpicture}
}
\caption{Implementation of LBFSF controller.}
\label{fig:LQR_blk_diag}
\end{figure}

Let $x_k^*$ and $u_k^*$ denote the optimal states and inputs. 
For all $k \in \{0, \ldots, N\},$ define
\begin{align*}
    A_k &\isdef \frac{\partial}{\partial x} f(x, u)\biggl|_{(x_k^*, u_k^*)} \in \BBR^{\ell_x \times \ell_x}, \\ 
    B_k &\isdef \frac{\partial}{\partial u} f(x, u)\biggl|_{(x_k^*, u_k^*)} \in \BBR^{\ell_x \times \ell_u}.
\end{align*}
%
%
Then, a full state feedback controller for reference tracking along the optimal trajectory derived in Section \ref{sec:NLP} is given by
\begin{equation}
    u_{\rm fb}(t) = K(t) (x^*(t) - x(t)),\label{eq:u_fb_opt}
\end{equation}
where the continuous time signals are reconstructed using zero-order hold, that is, $s(t) = s_k $ for all $t \in [k T_\rms, (k+1) T_\rms).$
At step $k$, the gain matrix $K_k$ is chosen such that $A_k + B_k K_k $ is Hurwitz. 
The implementation of the LBFSF controller is shown in Figure \ref{fig:LQR_blk_diag}.

\subsubsection{Autoregressive Moving Average (ARMA) Controller} \label{sec:ARMA}

\begin{figure}[h]
\centering
%
{%
    \begin{tikzpicture}[>={stealth'}, line width = 0.25mm]

    \node [input, name=ref]{};
    \node [smallblock, rounded corners, right = 0.5em of ref , minimum height = 2em, minimum width = 5em] (system) {System};
    \node [sum, left = 2em of system] (sum1) {};
    \node[draw = none] at (sum1.center) {$+$};
    \node [sum, right = 2em of system] (sum2) {};
    \node[draw = none] at (sum2.center) {$+$};
    \node [smallblock, fill=red!20, rounded corners, below = 1 em of system, minimum height = 2em , minimum width = 5em] (controller) {$\begin{array}{c} \mbox{ARMA} \\ \mbox{Controller} \end{array}$};
    \draw [->] (controller.west) -| node [above, near start] {$u_{\rm fb}$} (sum1.south);
    \draw[->] (system.east) -- node [above]{$y$} (sum2.west);
    \draw[->] (sum2.south) |- node [above, near end] {$y_{\rm fb}$} (controller.east);
    \draw[->] (sum1.east) -- node [above] {$u$} (system.west);
    \draw[->] ([xshift = -2em]sum1.west) -- node [above] {$u^*$} (sum1.west);
    \draw[<-] (sum2.east) -- node [above] {$y^*$} node [below, xshift = -0.5em] {$-$} ([xshift = 2em]sum2.east);
    \end{tikzpicture}
}
\caption{Implementation of ARMA controller.}
\label{fig:ARMA_blk_diag}
\end{figure}
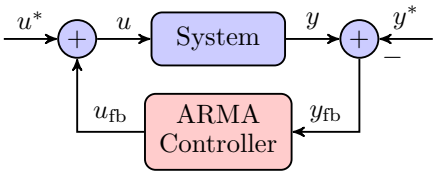

%
An ARMA controller can be written as
\begin{align}
    u_k = \phi_k \theta,
\end{align}
where 
\begin{align}
    \phi_k = \matl u_{k-1} & \ldots u_{k-l_\rmw} & y_{k-1} & \ldots y_{k-l_\rmw} \matr
\end{align}
and $\theta \in \BBR^{2 l_\rmw}$ is the vector of controller coefficients. 
Note that the ARMA controller is an $l_\rmw$th order linear controller driven by the output $y.$
ARMA controllers are described in more detail in Section IV.A of \cite{paredes2024mpc}.
The optimization of the ARMA controller coefficients is described in detail in Section V of \cite{paredes2024mpc}.
Figure \ref{fig:ARMA_blk_diag} shows the implementation of the ARMA controller.
In this work, we generate the training data for the ARMA controller coefficients optimization with the LBFSF controller, described in the previous section. 
%
%


%

\subsection{Fuzzy Control Systems} \label{sec:fuzzy}
A fuzzy control system consists of multiple controllers whose outputs are combined using fuzzy logic to compute the output of the fuzzy control system. 
In this work, we use the T-S fuzzy system framework \cite[ch.~6]{lilly2011}  to improve the robustness of the ARMA controller.
The implementation of the fuzzy system framework to combine ARMA controllers is described in Section IV.B in \cite{paredes2024mpc}.
In particular, the fuzzy controller's output, which consists of two controllers, is given by
\begin{equation}
    u_{\rm fuzzy}(t) 
        =
            \frac
            {\mu_{\rm 1} (\gamma) \ u_{\rm 1}(t)   + \mu_{\rm 2} (\gamma) \ u_{\rm 2}(t) }
            {\mu_{\rm 1} (\gamma) + \mu_{\rm 2} (\gamma)},
    \label{eq:fuzzy_controller}
\end{equation}
where $u_1(t)$ and $u_2(t)$ are the outputs of the first and second controller and $\mu_1(t)$ and $\mu_2(t)$ are the corresponding membership functions.

\section{Simulation Results} \label{sec:numerical_examples}

In this section, we solve the NLP in Section \ref{sec:NLP} to obtain the time-optimal state and input trajectory of a bicopter in a maneuver that consists of a vertical displacement and a flip.
The time-optimal trajectories are then used to construct the feedback controllers described in Section \ref{sec:fbc}.
The performance of the feedback controller is demonstrated in numerical simulations.
In this work, we use the Simulink environment to simulate the bicopter and the controller.
%


\subsection{Time Optimal Trajectory}
\label{subsec:bicopter_traj_optim}

To compute the time-optimal trajectory, we solve the nonlinear program described in Section \ref{sec:NLP}.
The bicopter is commanded to start from the initial state $x_\rmi$ and reach the final state $x_\rmf,$ where 
\begin{align}
    x_\rmi
        &\isdef
            \begin{bmatrix} r_1(0) & \dot{r}_1(0) & r_2(0) & \dot{r}_2(0) & \psi(0) & \dot{\psi}(0) \end{bmatrix}^\rmT 
        \nn \\
        &= 
            0,
    \\
    x_\rmf 
        &\isdef
            \begin{bmatrix} r_1(T) & \dot{r}_1(T) & r_2(T) & \dot{r}_2(T) & \psi(T) & \dot{\psi}(T)\end{bmatrix}^\rmT \nn \\
        &=
            \begin{bmatrix} 0 & 0 & 3 & 0 & 2 \pi & 0 \end{bmatrix}^\rmT,
\end{align}
and the time $T$ is to be minimized. 
Furthemore, the final input conditions are
\begin{equation}
u_\rmf = \begin{bmatrix} u_\rmT(T) & u_\rmR(T) \end{bmatrix}^\rmT = \begin{bmatrix} g & 0 \end{bmatrix}^\rmT,
\end{equation}
Note that the initial- and final-state specifications encode the objective of moving 3 m upward and performing a flip in the middle of the trajectory.
Note that these conditions do not explicitly state the time at which the bicopter needs to perform the flip.
Instead, the maneuver is determined by the optimization algorithm.
%

\begin{figure}[h]
    \centering
    \includegraphics[width=0.60\columnwidth]{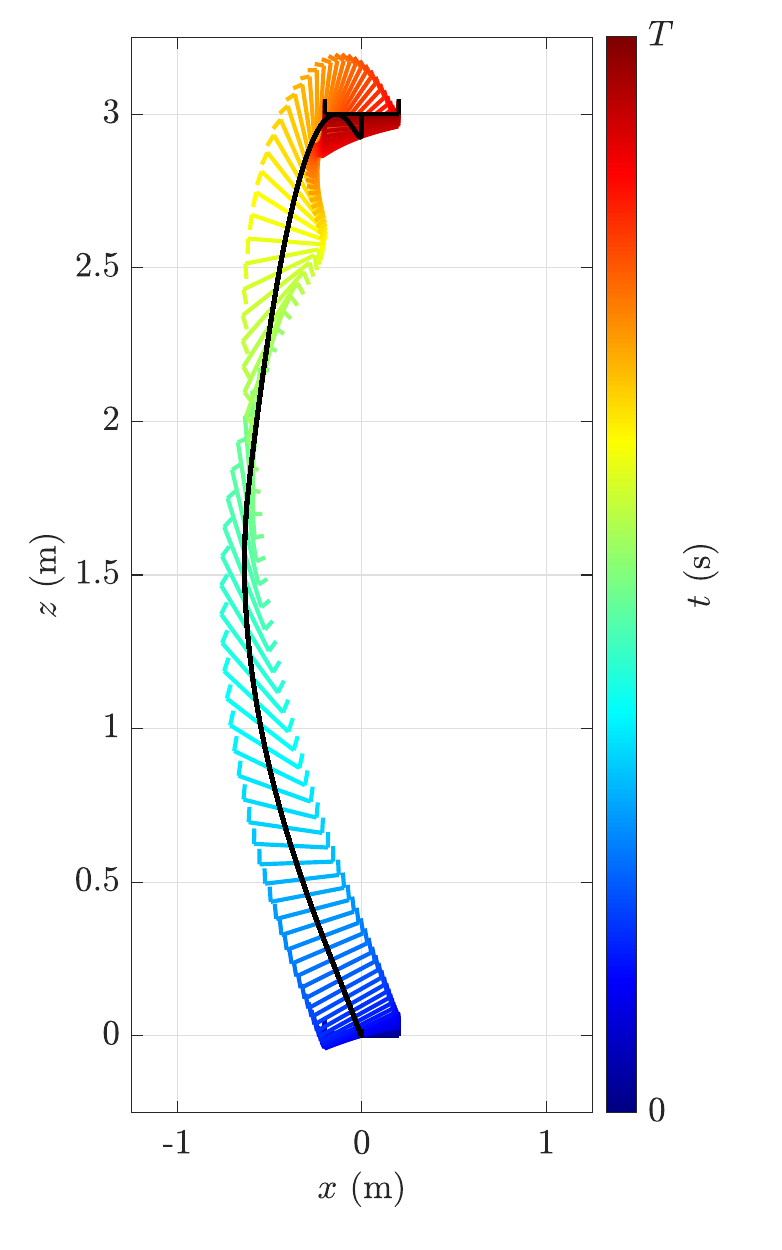}
    \caption{Optimal trajectory of the bicopter with a flip maneuver computed by solving a nonlinear program using CasADi. }
    \label{fig:MC_traj_optim_vertical_2D}
\end{figure}

\begin{figure}[h]
    \centering
    \includegraphics[width=0.95\columnwidth]{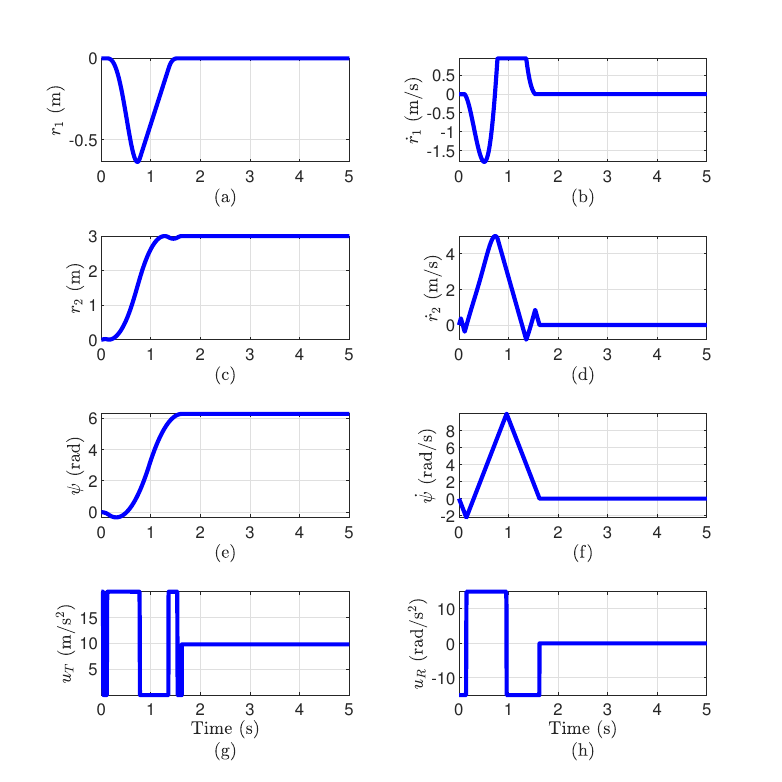}
    \caption{Bicopter states and inputs in the optimal trajectory.
    }
    \label{fig:optimalTrajectory}
\end{figure}

%
%
Next, the input constraints are set as 
\begin{align}
    u_{k, (1)} &= u_{\rmT, k} 
    \in [1, \ 20] \mbox{ m/s$^2$},\\
    u_{k, (2)} &= u_{\rmR, k} 
    \in [-15, \ 15] \mbox{ rad/s$^2$},
\end{align}
state constraints are set as
\begin{align}
    x_{k, (1)} &= r_{k,(1)} \in [-1, \ 1] \mbox{ m},\\
    x_{k, (3)} &= r_{k,(2)} \in [0, \ 3] \mbox{ m}.
\end{align}
Note that the state constraints are set to prevent the optimization algorithm from yielding trajectories that are too far away from the line trajectory between the initial and final state conditions.
These constraints may be relaxed to improve optimization convergence.
The optimal trajectory is computed by solving the nonlinear program (NLP) given by \eqref{eq:NLP_init}-\eqref{eq:NLP_fin} using the CasADi Opti stack \cite{andersson2019} with the IPOPT numerical solver with $N = 400,$ $Q_x = 100 I_6,$ and $Q_u = 100 I_u.$ 
The initial guess for the final optimal time is set to $T = 10$ s to improve the optimization procedure, and the initial guess for the rest of the optimization parameters is set to zero.
%
Figure \ref{fig:MC_traj_optim_vertical_2D} shows the optimal trajectory of the resulting bicopter with a flip maneuver.
Figure \ref{fig:optimalTrajectory} shows the bicopter states and inputs in the optimal trajectory.
%
%
The optimized trajectory had a final time of $T = 1.6278$ s.
%
%

\subsection{Effect of Initial Condition Perturbations}
\begin{figure}[h]
    \centering
    \includegraphics[width=0.95\columnwidth]{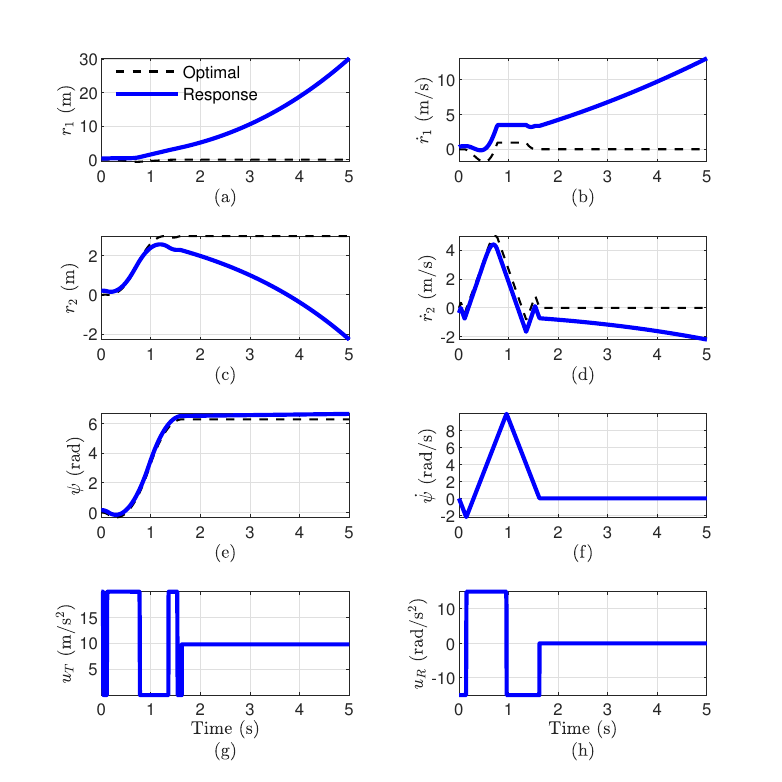}
    \caption{Bicopter states and inputs in the case where the optimal input trajectory is applied and initial conditions are given by \eqref{eq:x0_1}.
    %
    %
    %
    }
    \label{fig:openLoopResponse}
\end{figure}

In practice, the generated optimal input trajectory can not be applied to the physical system since any variation in the initial conditions or the physical properties of the system may result in deviations from the generated optimal state trajectory. 
In this subsection, we consider the application of the optimal input trajectory with nonzero initial conditions.
In particular, we consider the initial condition given by
%
%
\begin{align}
    x(0) = \matl 0.32 &
    0.32 &
    0.22 &
    -0.35 &
    0.16 & 
    0.02 \matr^\rmT, \label{eq:x0_1}
\end{align}
which is obtained by sampling from a uniform distribution between $-0.5$ and $0.5$.
Figure \ref{fig:openLoopResponse} shows the response of the bicopter under the optimal input trajectory and the initial conditions \eqref{eq:x0_1}, where the optimal state and input trajectories are shown in black dashes and the system response is shown in solid blue lines. 
Note that the bicopter position and angle responses diverge from the optimal state trajectory for all $t > T \approx 1.6$ s, since the bicopter velocity responses obtained by applying the optimal input sequence do not converge to zero. 

\subsection{Feedback Control}

As shown in the previous section, the optimal control does not yield the optimal state response if there are any perturbations in the initial conditions.
In other words, the optimal control must be computed for each set of initial conditions, which is intractable. 
To follow the optimal state response, we consider the control law given by \eqref{eq:u_fb}.

\subsubsection{LBFSF Control.}
We first consider the LBFSF controller given by \eqref{eq:u_fb_opt},
where $x(t)$ is the state of the bicopter,  $x^*(t)$ is the optimal state computed by CasADi, and the gain $K(t)$ is computed using the Matlab LQR routine.
%
The chosen LQR controller design parameters are set as $R_1 = I_6, R_2 = I_2.$
%
%

\begin{figure}[h]
    \centering
    \includegraphics[width=0.95\columnwidth]{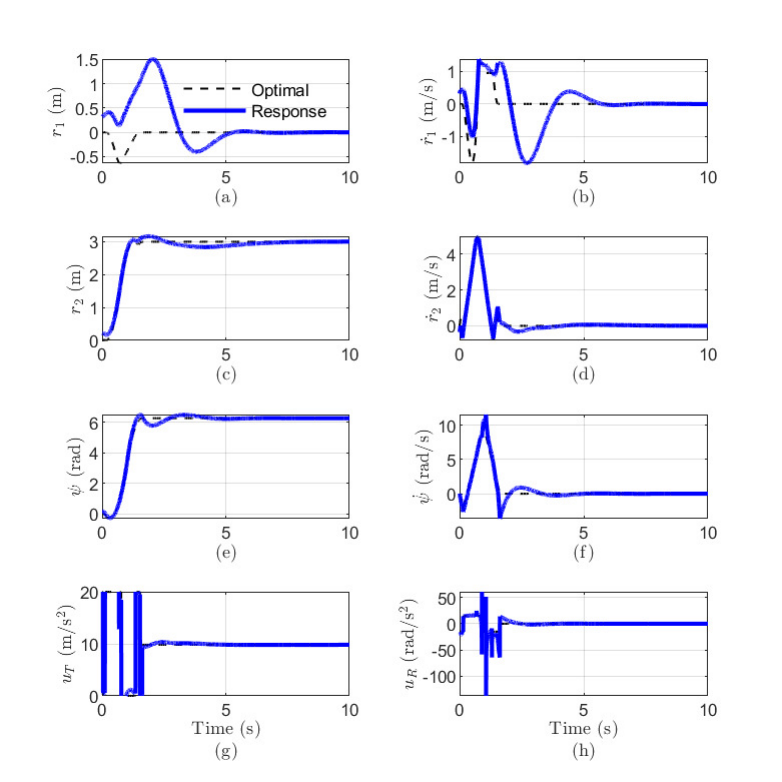}
    \caption{ Bicopter states and inputs in the case where the input is given by the LBFSF controller \eqref{eq:u_fb_opt}, where $K$ is obtained at each simulation  step, and the initial conditions are given by \eqref{eq:x0_1}.
    %
    }
    \label{fig:FBLQR}
\end{figure}

%
Figure \ref{fig:FBLQR} shows the response of the bicopter with the LBFSF controller. 
Note that the control signal $u_{\rm fb}(t)$ is able to compensate for the unknown initial conditions and follow the optimal state trajectory. 

\subsubsection{Learning-based ARMA Control}
Although the LBFSF controller developed in the previous section performs well, it is computationally expensive and requires fullstate feedback. 
In this section, we therefore design a fixed-gain ARMA controller.
To learn the ARMA controller coefficients, we generate 100 trajectories with the linearization-based control with random initial conditions. 
From each trajectory, the input sequence $u_{\rm fb}(t) \in \BBR^2$ and the output sequence 
\begin{align}
y_{\rm fb}(t) \isdef \matl r_1(t) - r_1^*(t) \\ r_2(t) - r_2^*(t) \\ \psi(t) - \psi^*(t) \matr    \in \BBR^3
\end{align}
are obtained at $t = k T_\rms,$ where $T_\rms = 10^{-3}$ second is the sampling rate of the controller. 

Using the controller learning process based on the least-square regression described in \cite{paredes2024mpc},  we construct two decoupled controllers. 
The first controller is driven by the position error, and the second controller is driven by the angular error.
%
In this work, we set $l_\rmw = 5$ for both controllers.  

Note that the discrete-time control signal generated by the ARMA controller at each discrete-time control iteration $k$ is given by $u_{{\rm arma}, k}.$
The continuous-time control signal $u_{\rm fb}(t)$ applied to the system is generated by applying a zero-order-hold operation to $u_{{\rm fb}, k},$ that is,
for all $k\ge0,$ and, for all $t\in[kT_\rms, (k+1) T_\rms),$ 
$u_{\rm fb}(t) = u_{{\rm arma}, k}.$

\begin{figure}[h]
    \centering
    \includegraphics[width=0.95\columnwidth]{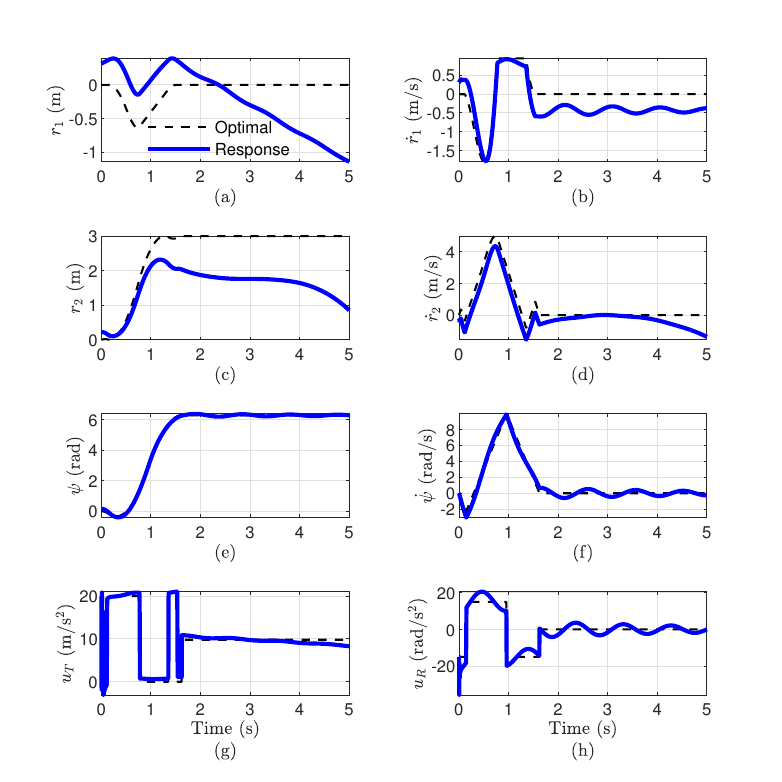}
    \caption{Bicopter states and inputs in the case where the input is given by the ARMA controller and the initial conditions are given by \eqref{eq:x0_1}.
    %
    }
    \label{fig:ARMA}
\end{figure}

Figure \ref{fig:ARMA} shows the response of the bicopter with the trained ARMA controller.
Note that while the controller can follow the optimal state trajectory, the bicopter deviates from hover conditions after the end of the optimal state trajectory.


\subsubsection{Fuzzy Controller}


As shown in the previous two case studies, the LBFSF controller provides good performance but is computationally expensive due to the need to compute the linearized matrices and the corresponding controller gains at each time instant, whereas the learned ARMA controller is computationally inexpensive but suffers from poor performance near the hover conditions.
To combine the advantages of both controllers, we construct a fuzzy controller that consists of the LBFSF control, which is active near hover conditions, and the learned ARMA controller, which is active during aggressive maneuvers. 
Furthermore, the LBFSF controller is obtained only at the hover conditions, that is, the controller gains are computed offline at the hover condition, thus reducing the computational cost of the controller; 
note that under these conditions the LBFSF controller is equivalent to a hover-condition linearized LQR controller.

%

\begin{figure}[h!]
    \centering
    \includegraphics[width=0.95\columnwidth]{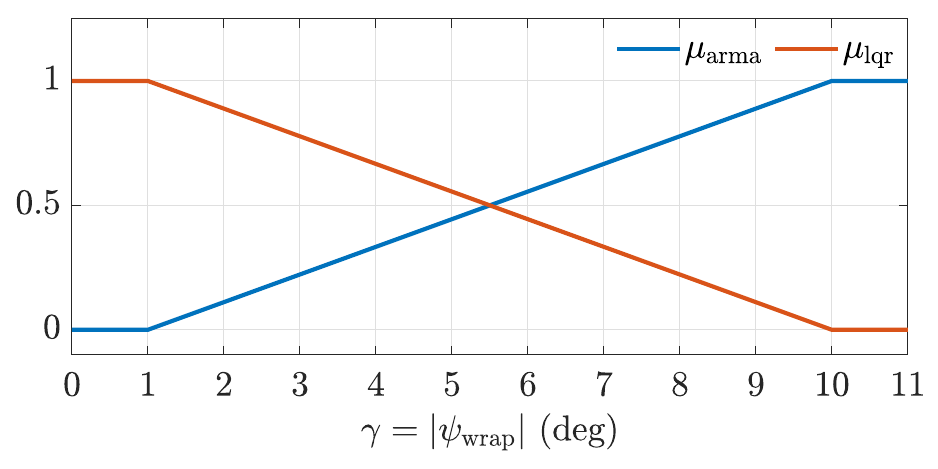}
    \caption{Fuzzy membership functions $\mu_{\rm arma}$ and $\mu_{\rm lqr}$ given $\gamma = |\psi_{\rm wrap}|,$ where $\psi_{\rm wrap} \in [-180, 180]$ deg is the wrapped angle $\psi$.} 
    \label{fig:mu_fuzzy}
\end{figure}

In this work, we use the T-S fuzzy system to combine the two controllers. 
Defining $\gamma \isdef |\psi_{\rm wrap}|,$ where $\psi_{\rm wrap} \in [-180, 180]$ deg is the wrapped angle $\psi,$ and letting $u_{\rm arma}, u_{\rm lqr}$ be the control signals generated by the ARMA and LQR controllers, the control signal generated by the fuzzy system is
\begin{equation}
    u_{\rm fuzzy}(t) 
        =
            \frac
            {\mu_{\rm arma} (\gamma) \ u_{\rm arma}(t)   + \mu_{\rm lqr} (\gamma) \ u_{\rm lqr}(t) }
            {\mu_{\rm arma} (\gamma) + \mu_{\rm lqr} (\gamma)}, \label{eq:fuzzy_arma_lqr}
\end{equation}
where $\mu_{\rm arma}$ and $\mu_{\rm lqr}$ are the fuzzy membership functions associated with the ARMA and LQR controllers, respectively, shown in Figure \ref{fig:mu_fuzzy}.
%

\begin{figure}[h]
    \centering
    \includegraphics[width=0.95\columnwidth]{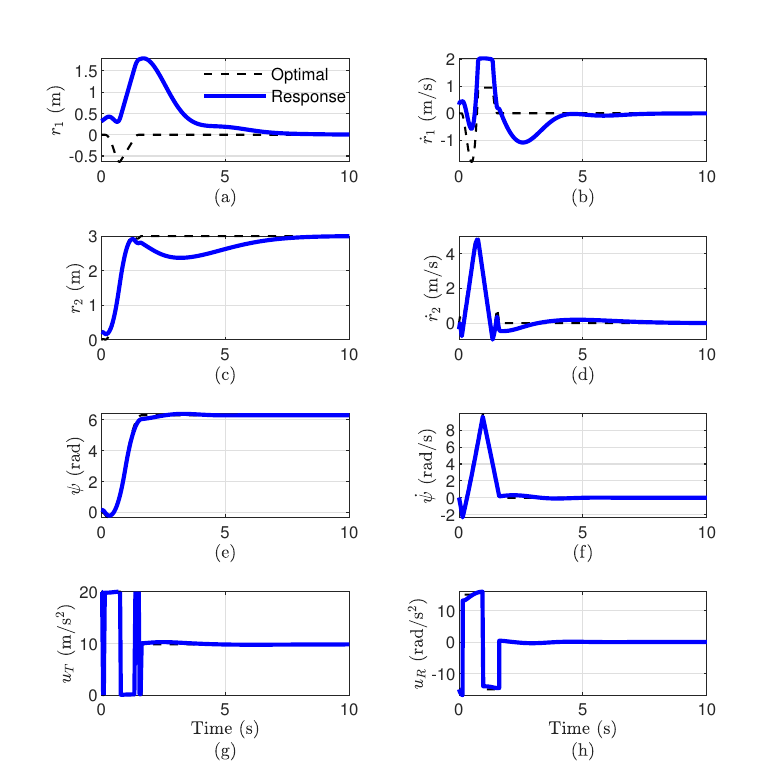}
    \caption{
    Bicopter states and inputs in the case where the input is given by the fuzzy controller shown in \eqref{eq:fuzzy_arma_lqr}, which combines the trained ARMA controller and a hover-condition linearized LQR controller, and the initial conditions are given by \eqref{eq:x0_1}.
    %
    }
    \label{fig:F_ARMA_LQR}
\end{figure}

%
%
Figure \ref{fig:F_ARMA_LQR} shows the response of the bicopter with the fuzzy controller obtained from combining the responses of the trained ARMA controller and the hover-condition linearized LQR controller.
%
%
Note that the fuzzy controller's closed-loop performance is similar to that of the linearization-based control, but at a significantly reduced computational cost since all of the controller gains of the fuzzy controller are fixed. 


%
%
%


\section{Conclusions} \label{sec:conclusions}
This paper presented a data-driven fuzzy controller synthesis framework for time-optimal trajectory following.
In this framework, the time-optimal trajectory for an aggressive trajectory is obtained using a numerical solver, and an ARMA controller is trained to mimic the obtained trajectory.
The responses of the trained ARMA controller and a stabilizing controller then weighted depending on the measured system conditions and interpolated using a T-S fuzzy system.
A multicopter numerical example illustrates the efficiency of this framework, and shows that the fuzzy controller is able to track the time-optimal trajectory. 

%

\bibliography{bib_paper.bib,Bib/PX4bib}

\end{document}